\documentclass[aps,showpacs,floats,prb,amsmath,amssymb,twocolumn]{revtex4}
\usepackage[applemac]{inputenc}
\usepackage{graphicx}

\def\be{\begin{equation}}
\def\ee{\end{equation}}

\def\bi{\begin{itemize}}
\def\ei{\end{itemize}}
\def\bn{\begin{enumerate}}
\def\en{\end{enumerate}}
\def\bea{\begin{eqnarray}}
\def\eea{\end{eqnarray}}
\def\no{\nonumber}
\def\ba{\begin{array}}
\def\ea{\end{array}}
\def\bd{\begin{displaymath}}
\def\ed{\end{displaymath}}

\def\la{\langle}
\def\ra{\rangle}

\def\gsim{\mathrel{\rlap{\lower4pt\hbox{\hskip1pt$\sim$}}
\raise1pt\hbox{$>$}}}  

\newcommand{\bk}{{\bf k}}

\newcommand{\bQ}{{\bf Q}}

\newcommand{\barg}{\bar{\gamma}}

\newcommand{\LiX}{Li$_{\text 2}$VO$X$O$_{\text 4}$}

\newcommand{\AAVO}{$AA'$VO(PO$_{\text 4}$)$_{\text 2}$}

\begin{document}

\title{Quantum fluctuations in high field magnetization of 2D square
lattice J$_1$-J$_2$ antiferromagnets.}

\author{P. Thalmeier$^1$, M. E. Zhitomirsky$^2$, B. Schmidt$^1$, N. Shannon$^3$}
\affiliation{$^1$Max Planck Institute for Chemical Physcis of Solids, 01187 Dresden, Germany\\
$^2$ Commissariat \'a l'Energie Atomique, DSM/DRFCM/SPSMS, 38054 Grenoble, France\\
$^3$ H H Wills Physics Laboratory, Bristol BS8 1TL, United Kingdom}
%

\date{\today}
\begin{abstract}
\leftskip 2cm \rightskip 2cm
The $J_1$--$J_2$ square lattice Heisenberg model with spin
$S=1/2$ has three phases with long-range magnetic order
and two unconventionally ordered phases
depending on the ratio of exchange constants. It describes a number
of recently found layered vanadium oxide compounds. A simple means
of investigating the ground state is the study of the magnetization
curve and high-field susceptibility. We discuss these quantities by
using the spin-wave theory and the exact diagonalization in 
the whole $J_1$--$J_2$ plane. We compare both results and
find good overall agreement in the sectors of the phase diagram with magnetic order. 
Close to the disordered regions the magnetization curve shows strong deviations 
from the classical linear behaviour caused by large quantum fluctuations and spin-wave 
approximation breaks down.
On the FM side ($J_1<0$)  where one approaches the quantum gapless spin nematic ground state  
this region is surprisingly large. We find that inclusion of second order spin-wave corrections 
does not lead to fundamental improvement. Quantum corrections to the tilting angle of 
the ordered moments are also calculated. They may have both signs, contrary to  the always 
negative first order quantum corrections to the magnetization. Finally we investigate 
the effect of the interlayer coupling and find that the quasi-2D picture remains 
valid up to $|J_\perp/J_1|\sim 0.3$.
\end{abstract}

\maketitle

\section{Introduction}


The search for a quantum spin-liquid --- an insulating magnet with a gapless ground state which breaks neither 
lattice nor spin symmetries  --- has focused largely on spin-1/2 two-dimensional quantum antiferromagnets (2DQAF's).   
In practice, however, most two-dimensional AF spin-1/2 Heisenberg models exhibit either N\'eel order, or crystals of short-ranged 
singlet bonds and a finite gap to spin excitations~\cite{Misguich04}.   In a few cases, gapless {\it hidden order} states
with nematic character arise~\cite{Laeuchli05}, but among ``realistic'' models perhaps only the S=1/2 
Heisenberg model on a Kagom\'e lattice remains a serious candidate for a spin-liquid description 
(see, e.g.~\cite{Misguich07} and references therein).

Perhaps the best studied example of a 2DQAF is the spin-1/2 $J_1$--$J_2$ Heisenberg 
model,  which demonstrates a quantum phase transition from N\'eel to valance bond solid
as a function of the control parameter $J_2/J_1$~\cite{Misguich04}.  A number of 
layered vanadium 
compounds have recently been synthesized which are well described 
by this model.   They are of the type  \LiX{} ($X=\text{Si},\text{Ge}$)~\cite{Millet98, Melzi00, Melzi01} and \AAVO{}
($A,A'=\text{Pb},\text{Zn},\text{Sr},\text{Ba}$)~\cite{Kaul04,Kaul05,Kini06}
consisting of vandium oxide pyramid layers containing  V$^{4+}$ ions with spin $S=1/2$. 

In Refs.~\onlinecite{Shannon04,Schmidt07} an extensive analysis of the $J_1$--$J_2$ model, 
also for finite magnetic field has been given in order to understand physical properties of  
the above compounds. Both the numerical exact diagonalization (ED) Lanczos method for finite clusters and 
the analytical spin-wave analysis have been employed. The behaviour of the saturation field as a function 
of the frustration angle has been studied as a further means of diagnosis of $J_1$--$J_2$ compounds. 
It was found that close to the disordered regime with $J_1<$ 0 it is determined by an instability of 
two spin excitations, indicating indeed that in this regime the ground state may be of a 
spin-nematic type. \cite{Shannon06} This leads us to expect that the magnetization itself should
also be anomalous in this regime with a large effect from quantum fluctuations. So far 
quantum corrections to the magnetization curve in the spin-wave theory have 
only been considered for the nonfrustrated square-lattice antiferromagnet 
($J_2$=0). \cite{Zhitomirsky98} 
 
In this work we give a systematic investigation of the magnetization and the high-field 
susceptibility  for the general  2D square lattice $J_1$--$J_2$ model. Our goal is 
to investigate how the quantum corrections on these quantities depend on the degree of frustration, 
especially close to the disordered phases.  The effect of interplane coupling will be also 
considered.  

The low scale of exchange interactions which are of order 10 K in the \LiX{}  and and \AAVO{} vanadates mean 
that magnetization measurements are relatively easy to perform.   
It is therefore hoped that the analysis with the theory developed  here gives an additional criterion
to determine the frustration ratio $J_2$/$J_1$ in a specific compound. 
In Sec.~\ref{sect:J1J2model} we briefly introduce the model. Its high field properties 
like magnetization and susceptibility are investigated in Sec.~\ref{sect:HIGHFIELD}. They are 
obtained from ED Lanczos calculations  and  first as well as second order spin-wave theory 
and a comparison is given. We also evaluate the contributions of interlayer coupling 
in Sec.~\ref{sect:interlayer} and in Sec.~\ref{sect:DISC} we finally give a discussion and conclusion.

\section{The J$_1$-J$_2$ model and its phases}
\label{sect:J1J2model}

The 2D square lattice spin-1/2 Heisenberg model in an
external magnetic field H  is given by 
\begin{eqnarray}
\label{eq:HAMJ1J2}
{\cal H} &=& J_1\sum_{\langle ij \rangle_1} {\bf S}_i.{\bf S}_j 
 + J_2\sum_{\langle ij \rangle_2} {\bf S}_i.{\bf S}_j
 - h \sum_i S^z_i \ ,
 \label{EXHAM}
\end{eqnarray}
where $J_1$ and $J_2$ are the two exchange constants between the first and 
the second neighbors on a square lattice, respectively.
As in Refs.~\onlinecite{Shannon04,Schmidt07} the exchange parameters are defined 
per exchange bond. Furthermore we use the convention $h=g\mu_B H$ ($g$ = gyromagnetic
ratio, $\mu_B$ = Bohr magneton). The phases in zero field are best
characterized by introducing equivalent parameters
\begin{eqnarray}
J_{\text c}=(J^2_1+J^2_2)^\frac{1}{2}, \quad \phi=\tan^{-1}(J_2/J_1) \ ,
\label{SCALE}
\end{eqnarray}
or $j=\tan\phi = J_2/J_1$. The angle $\phi$ determines the extent of
magnetic frustration in the model. 

Three classical magnetic ground
states are possible depending on $\phi$; namely ferromagnet (FM),
N\'eel antiferromagnet and collinear antiferromagnet (CAF). They have
been extensively discussed in Ref.~\onlinecite{Shannon04}. The effect
of exchange frustration leading to enhanced quantum fluctuations is
strongest at the classical phase boundaries where the CAF phase joins
the NAF or FM phase (see insets of Fig.~\ref{fig:mu_ed}). In
fact in these regions they are believed to destroy long-range magnetic order
and establish two new partially ordered states, namely a columnar dimer state with a
spin gap at CAF/NAF boundary \cite{Gelfand89,Sachdev90,Schulz92,Kotov99,Sirker07} 
and a gapless spin nematic
state at the CAF/FM boundary \cite{Shannon06}. It can be seen
already within spin wave approximation that the magnetic order breaks
down in this regime since the sublattice moment reduction due to
quantum fluctuations diverges close to the two boundary regions
\cite{Shannon04}.

\section{High field properties of the J$_1$-J$_2$ model}
\label{sect:HIGHFIELD}

The determination of the frustration ratio $J_2/J_1$ or angle
$\phi=\tan^{-1}(J_2/J_1)$ is of foremost importance to characterize a
given square lattice magnetic compound such as the SiO$_4$ and PO$_4$
vanadates mentioned in the introduction.  The available experimental
methods to reach this objective have been extensively discussed in
Ref.  \onlinecite{Shannon04}.  However, the thermodynamic zero-field
methods like evaluating the heat capacity and the magnetic
susceptibility are too ambiguous to locate $\phi$ in the NAF or CAF
sector of the phase diagram.  Additional information may be obtained
from investigating high-field properties \cite{Schmidt07}, for example
from saturation fields as, e.g., determined from the magnetocaloric
effect.  

The analysis of the high field magnetization itself as
function of frustration angle $\phi$ is also promising.  From the
simple n.n. Heisenberg AF ($J_2=0$ or $\phi=0$) it is known from
analytical work \cite{Zhitomirsky98} that deviations from the
classical linear magnetization curve due to quantum fluctuations are
to be expected.  This is also concluded from numerical calculations.
\cite{Yang97}  A systematic study of magnetization curves
for the $J_1$--$J_2$ model is however lacking.  Since the method is of
experimental importance due to its relative simplicity we consider it
worthwile to investigate this problem in detail for the
antiferromagnetic phases of the $J_1$--$J_2$ model.  For this purpose
we will use both analytical spin wave methods similar to
Ref.~\onlinecite{Zhitomirsky98} and numerical Lanczos methods for
exact diagonalization of finite clusters.  Our notation will be close
to the one used in Ref.~\onlinecite{Schmidt07}.

\subsection{Magnetization from numerical ($T=0$) Lanczos results for
$J_1$-$J_2$ clusters}
\label{sect:LANCZOS}
%
\begin{figure*}
    \centering
    \hfill
    \includegraphics[width=.4\textwidth]{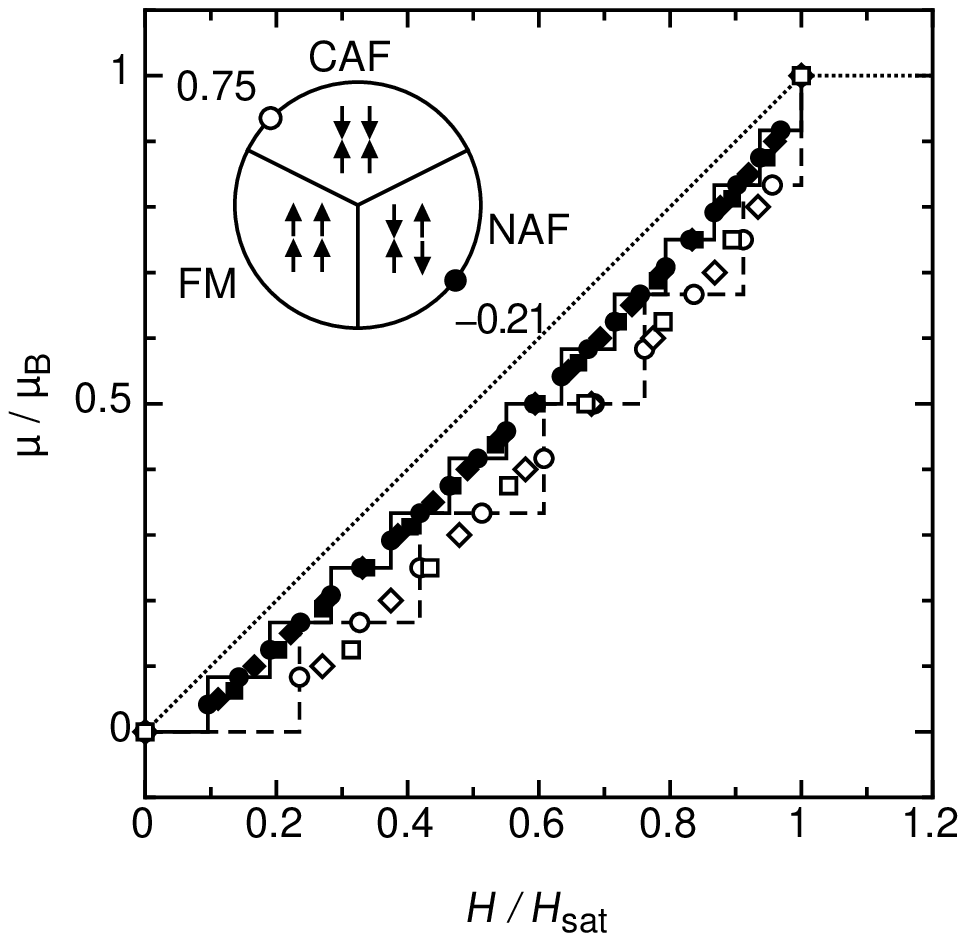}
    \hfill\hfill
    \includegraphics[width=.4\textwidth]{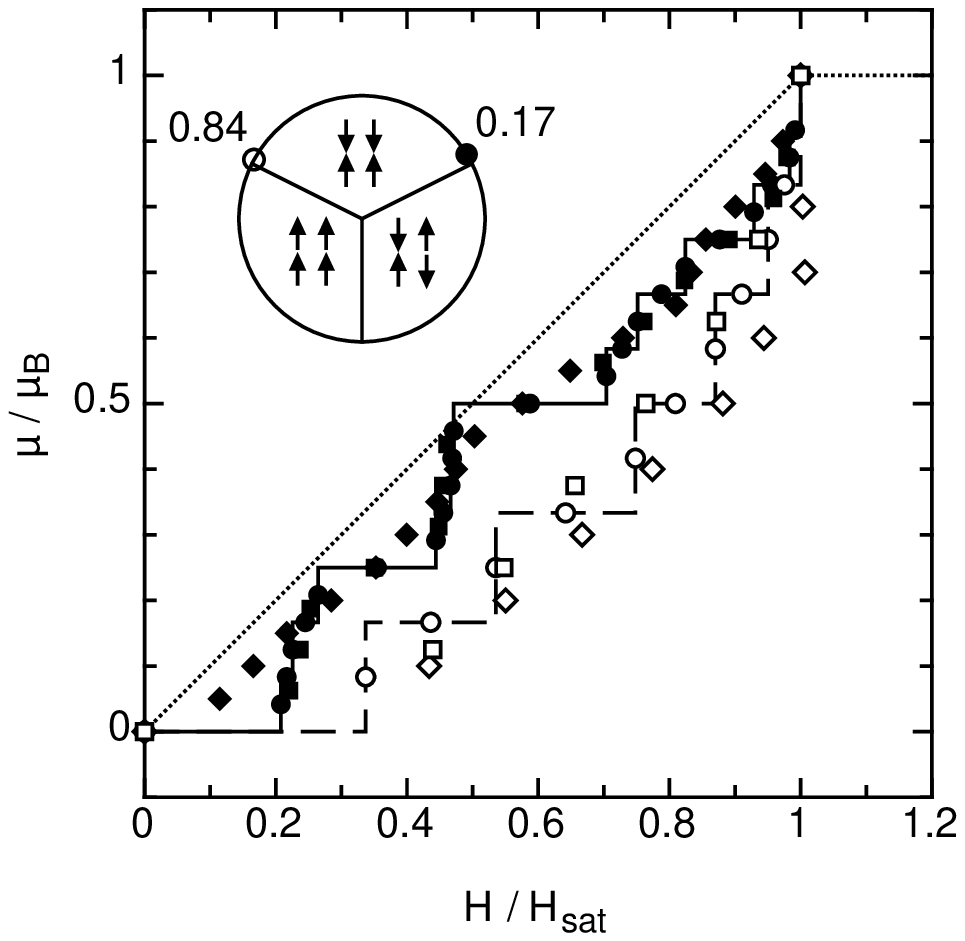}
    \hfill\null
    \caption{The normalized magnetization obtained from
     the Bonner-Fisher construction  
     as a function of an applied magnetic field
    $H$ in units of the saturation field $H_{\text{sat}}$ for
    different frustration angles (for $J_1>0$ H$_{sat}$ = $g\mu_Bh_s$).
    Data from the 24-site cluster are
    denoted by circles, 20-site data by diamonds, and 16-site data by
    squares.  Additionally, the zero-temperature magnetization steps
    are plotted for the 24-site clusters; the solid line corresponds
    to the solid symbols, the dashed line to the open symbols.  The
    insets in the plots show the positions of the frustration angles  $\phi/\pi$
    in the classical phase diagram. The frustration angle is counted positive above
    ($0<\phi/\pi < 1$) and negative below ($-1<\phi/\pi < 0$)  the x-axis in
    the inset diagram. For reference, the classical magnetization
    curve is plotted as a dotted line in both plots.  Left:
    Classically ordered phases.  N\'eel phase (NAF), $\phi=-0.21\,\pi$
    (solid symbols) and collinear phase (CAF), $\phi=0.75\,\pi$ (open
    symbols).  Right: Disordered regimes.  Columnar-dimer phase
    (crossover from NAF to CAF), $\phi=0.17\,\pi$ (solid symbols) and
    spin-nematic phase (crossover from CAF to FM), $\phi=0.84\,\pi$
    (open symbols).}
    \label{fig:mu_ed}
\end{figure*}
%
We have numerically diagonalized the Hamiltonian of the
$J_1$--$J_2$ model for cluster sizes of 16, 20, and 24 sites
applying the finite-temperature Lanczos method as described in
Ref.~\onlinecite{Schmidt07} and references cited therein.  In the
limit $T\to0$, our results are identical to the standard
zero-temperature implementation of the algorithm, i.\,e., the
evaluation of the partition function reduces to the determination of
ground-state expectation values.  The clusters of size 16 and 20 are
(regular and tilted) squares, the 24-site cluster is a rectangle.  All
three clusters tile the infinite lattice such that, with periodic
boundary conditions, compatibility with the three classically ordered
ground states is preserved.

The zero-temperature field dependence of
the magnetization $m=\langle S^z\rangle$ has been calculated 
for the whole phase diagram except the FM region, where
the saturation field $h_s$ vanishes.
Following
Ref.~\onlinecite{Bonner64}, Fig.~\ref{fig:mu_ed} shows 
the normalized magnetization $m/m_\text{sat}$ ($m_\text{sat}\equiv S$) for
selected values of the frustration angle $\phi$ in the classically
ordered antiferromagnetic phases and in the two disordered regimes of
the phase diagram.  The magnetic field is normalized to the saturation
field determined by exact diagonalization.  This is identical to the
classical saturation field for {\em positive\/} (antiferromagnetic) $J_{1}$.  For {\em
negative\/} (ferromagnetic) $J_{1}$, a $\Delta S=2$ two-magnon instability determines
the saturation field for the finite-size systems considered here,
which occurs at slightly higher field values than the one-magnon
instability.~\cite{Schmidt07}

The field dependence of the magnetization at $T=0$ for a finite-size
system is a sequence of finite steps.
The solid line in the left panel of
Fig.~\ref{fig:mu_ed} shows the magnetization curve of the 24-site
cluster for $\phi=-0.21\,\pi$, which is in the N\'eel phase, the dashed
line shows the same for $\phi=0.75\,\pi$ in the collinear phase.  In
the right panel, the solid line shows the field dependence for
$\phi/\pi=0.17$ (columnar dimer phase), the dashed line for
$\phi/\pi=0.84$ (spin-nematic phase).  The symbols in the plots denote
the midpoints of the horizontal and vertical line segments of the
magnetization steps; circles label the 24-site data, diamonds the
20-site data, and squares the 16-site data.  (For the smaller cluster
sizes, the step functions are not shown.)  Note that due to the
$\Delta S=2$ steps in the magnetization for $\phi>\pi/2$
(ferromagnetic $J_{1}$) there exist only half as much data points than
for $\phi<\pi/2$.  According to Bonner and Fisher (Ref.~\onlinecite{Bonner64}), 
these midpoints should yield, at least for $J_{2}=0$, a good approximation
of the magnetization curve of the square-lattice Heisenberg model in
the thermodynamic limit.

The angles $\phi/\pi=-0.21$ and $0.75$ for the NAF and CAF in the left
plot are chosen such that they correspond to the experimental findings
for the compound SrZnVO(PO$_{\text 4}$)$_{\text 2}$.~\cite{Shannon04}
For these values, the midpoints of the magnetization steps form a
smooth function.  This can generally be observed for any frustration
angle located inside the magnetically ordered regimes, be it NAF or
CAF. In contrast, for $\phi/\pi=0.17$ (solid symbols and line in the
right plot) and $\phi/\pi=0.85$ (open symbols and dashed line), the
data points scatter much more and do not give rise to a smooth field
dependence.  Furthermore, in the columnar dimer phase ($\phi/\pi=0.17$)
the well-stablished half-magnetization plateau that appears at $m=\frac{1}{2}m_\text{sat}$.
\cite{Zhitomirsky00,Honecker01}. The data in the right panel of Fig.~\ref{fig:mu_ed} 
indicate possibility of yet another plateau at $m=\frac{1}{4}m_\text{sat}$,
though a careful examination of fintie-size effects is necessary
to make a final conclusion.

In the whole phase diagram, quantum effects lead to negative corrections:
the exact magnetization curve $m(h)$ lies always below 
the corresponding classical value, which is a consequence 
of the lowering of the ground state energy of a quantum antiferromagnet
compared to its classical counterpart. \cite{Zhitomirsky98}
%
\begin{figure*}
    \includegraphics[width=.5\textwidth]{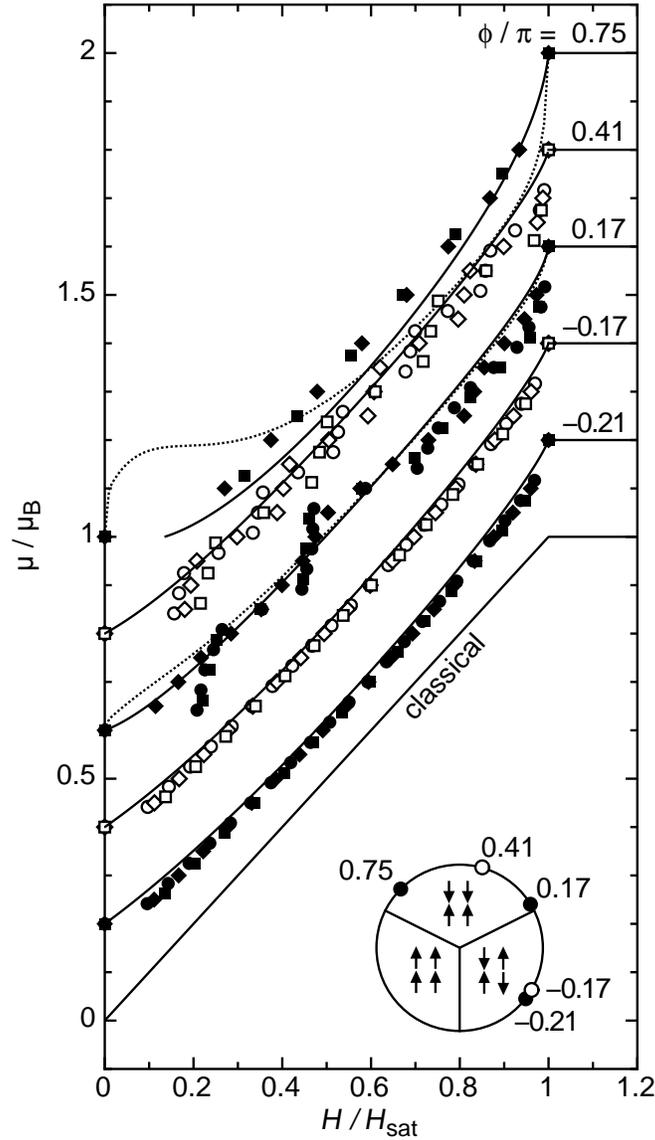}
     \vspace{1cm}
    \caption{Magnetization curves $\mu/\mu_B = gm$ $(=m/S)$ for
    various frustration angles in the antiferromagnetic or disordered
    sectors with an offset of 0.2 applied.  Symbols are obtained from Lanczos magnetization data
    for N = 16 (squares), 20 (diamonds), 24 (dots, circles) size
    clusters using the Bonner-Fisher construction \cite{Bonner64}.
    Lines are obtained from first (full) or second (dotted) order 
    spin wave calculations of Sects.~\ref{subsect:MAGNET1} and \ref{subsect:MAGNET2}.
    Angles $\phi/\pi$ = 0.75, -0.21 correspond to the possible CAF or
    NAF values of the Sr compound.  Magnetization curves strongly differ in the extent
    of nonlinear deviation from the classical curve which corresponds to $\phi/\pi$ = -0.5.
    Furthermore $\phi/\pi$ = 0.41, -0.17 are values which are deeply within the CAF
    or NAF regions and overall agreement of spin wave and Lanczos
    calculations is good.  It is less so on the CAF side where the
    magnetization changes in steps of $\Delta S_z$=2 leading to
    a larger finite size scattering.  The values $\phi/\pi$ = 0.75,
    0.17 correspond to regions close to or within the nonmagnetic sectors.
    Close to the CAF/FM boundary the first order spin wave results
    overemphasize the nonlinear behaviour and become unstable at very low
    fields. Close to the CAF/NAF boundary the numerical data exhibit a plateau at
    $m/S=\mu/\mu_B=0.5$ which would require a separate analysis.  The plateau was first
    reported in Ref.~\onlinecite{Zhitomirsky00}.  The inset shows the
    position of plotted $\phi$ values in the phase diagram. Second order spin wave results
    are discussed in  Sect.~\ref{subsect:MAGNET2}.}
    \label{fig:BonnerFisherZM1}
\end{figure*}
%

\subsection{Spin wave excitations in an external magnetic field}
\label{subsect:SW}

A standard Holstein-Primakoff approximation of
Eq.~(\ref{eq:HAMJ1J2}) and a subsequent Bogoliubov
transformation leads to the harmonic
spin wave Hamiltonian \cite{Schmidt07}
\begin{eqnarray}
    {\cal H} &=& NE_0  +NE_{zp} + \sum_{\lambda{\bf k}} 
    \epsilon_{\lambda\bk}(h)\alpha^{\dagger}_{\lambda{\bf k}} 
    \alpha^{\phantom\dagger}_{\lambda{\bf k}}\ ,
    \label{HPHAM}
\end{eqnarray}   
where $\alpha^{\dagger}_{\lambda{\bf k}}$ are magnon operators that
obey bosonic commutation rules. 
The  $\epsilon_{\lambda,\bk}(h)$ denote the spin wave dispersion of
branch $\lambda = \pm$ as defined
in the appropriate NAF or CAF magnetic Brillouin zone (BZ).  It is given by
\bea
\epsilon_{\pm\bk} (h) &=& \sqrt{[A_\bk(h) \pm C_\bk(h)]^2 -
B_\bk(h)^2} \ .
\label{DISP1}
\eea
Here $A_\bk(h)$ is the intra- and  $B_\bk(h)$, $C_\bk(h)$ are the
inter- sublattice couplings given below.
The ground state energy is composed of a classical part ($E_0$) 
obtained from the mean field approximation to Eq.~(\ref{eq:HAMJ1J2})
and a part due to zero point fluctuations of spins ($E_{zp}$). The
former is given by
\bea
\label{CLASSEN}
E_0&=&-h\la S_\parallel\ra+a_\parallel\la S_\parallel\ra^2-a_\perp\la
S_\perp\ra^2 \ ,
\eea
where $\la S_\parallel\ra =S\cos\frac{\theta_c}{2}$,  $\la S_\perp\ra
=S\sin\frac{\theta_c}{2}$ and $\theta_c/2$ is the classical field induced
canting angle of AF moments counted from the z-axis which is chosen
parallel to the applied field {\bf h} (here $\parallel \equiv z$ and $\perp\equiv x,y$). The
coefficients in Eq.~(\ref{CLASSEN}) are given by
\bea
a_\parallel &=& \frac{z}{2}(J_1+J_2)\ , \no\\
a_\perp &=& \frac{z}{2}(J_1-J_2) \ , \quad (\text{NAF}) \\
a_\perp &=& \frac{z}{2}J_2\ , \qquad \qquad (\text{CAF})  \no
\label{COEFFMF}
\eea
where $z = 4$ is the coordination number.
The classical canting angle $\theta_c/2$ of moments is obtained
minimizing E$_0$ which leads to
\bea
\cos\frac{\theta_c}{2}=\frac{h}{h_s} \ ,\quad h_s=2S(a_\parallel+a_\perp)\ ,
\label{CLASSANG}
\eea
where $h_s$ is the classical saturation field. For $h \geq h_s$
the moments are fully polarized, i.e., all are ferromagnetically
aligned parallel to {\bf h}. Explicitly we have $h_s = 2zSJ_1$ (NAF)
and $h_s = zS(J_1+2J_2)$ (CAF). 

The zero point energy due to quantum fluctuations is obtained as \cite{Schmidt07}
\begin{eqnarray}
 E_{\text{zp}}=\frac{1}{2N}\sum_{\lambda {\bf k}}\left[
\epsilon_{\lambda\bk}(h) - A_\bk\right]\ .
 \label{ZEROEN}
\end{eqnarray}   
For the CAF and NAF phases it is always
negative and vanishes in the FM or fully polarized (h $=$ h$_s$) phase
where the ground state and spin wave states are exact
eigenstates with a dispersion $\epsilon_{\lambda\bk} \equiv
A_\bk$. The sublattice couplings which determine the spin wave
dispersion in
the canted state were derived in Ref.~\onlinecite{Schmidt07} and are
given here for
completeness. Defining
\bea
 \label{COEFF1}
    A_\bk&=&Sa_\bk\ , \nonumber\\
    B_\bk(h)&=&Sb_\bk\sin^2\frac{\theta_c}{2}\ , \\
    C_\bk(h)&=&Sc_\bk\cos^2\frac{\theta_c}{2}\ ,  \nonumber
\eea
we have in the NAF and CAF phases corresponding to wave vectors
$\bQ=(\pi,\pi)$ and $ \bQ=(\pi,0)$, respectively:
\bea
\label{COEFF2}
a_\bk&=&4[J_1-J_2(1-\barg_\bk)] \ ,\quad (\text{NAF}) \no\\
c_\bk=-b_\bk&=&4J_1\gamma_\bk \ ,\\
a_\bk&=&2[2J_2+J_1\gamma_y] \ ,\quad\qquad (\text{CAF}) \no\\
c_\bk=-b_\bk&=&2[J_1+2J_2\gamma_y]\gamma_x  \ ,\no
\eea
where the geometric structure factors are defined by
\bea
\label{GSTRUC}
\gamma_\bk &=& \frac{1}{2}(\cos k_x +\cos k_y) \ , \no\\
\barg_\bk&=&\cos k_x\cos k_y \ , \\
\gamma_x &=& \cos k_x, \quad \gamma_y=\cos k_y \ . \no
\eea
Due to the symmetry properties $A_{\bk+\bQ} = A_{\bk}$,
$B_{\bk+\bQ}= -B_{\bk}$  
and $C_{\bk+\bQ} = -C_{\bk}$ we have the identity
$\epsilon_{\pm}(\bk+\bQ)=\epsilon_{\mp}(\bk)$.
Then, instead of summing over two spin wave branches in the
magnetic BZ of NAF or CAF, we may restrict to one
mode only, e.g. $\epsilon_\bk\equiv\epsilon_{+\bk}$ and sum over the
whole paramagnetic BZ in the expression for $E_{zp}$ and similar
ones. Using this convention the spin wave branch index $\lambda$ will be omitted
in the following . Using Eqs.~(\ref{DISP1}) and (\ref{COEFF1}) the spin
wave energies may be written as
\bea
\epsilon_\bk(h)=
 S(a_\bk+c_\bk)^\frac{1}{2}(a_\bk+c_\bk\cos\theta_c)^\frac{1}{2} \ ,
 \label{DISP2}
\eea
for both NAF and CAF cases with $a_\bk$ and $c_\bk$ given in
Eq.~(\ref{COEFF2}).

\subsection{Magnetization from first order spin wave quantum corrections}
\label{subsect:MAGNET1}
%
\begin{figure*}
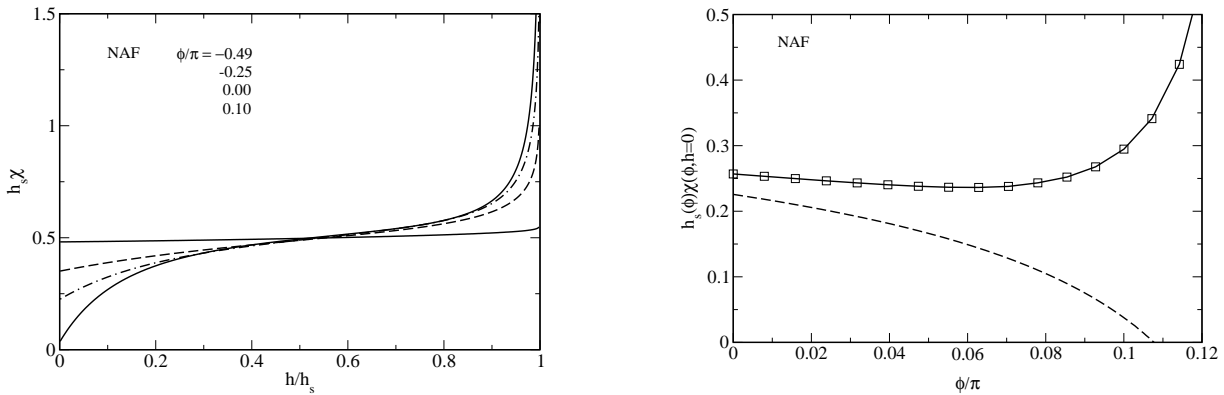

    \centering
    \hfill
    \includegraphics[width=.4\textwidth]{Fig3a}
    \hfill\hfill
    \includegraphics[width=.4\textwidth]{Fig3b}
    \hfill\null
    \caption{Left panel: Susceptibility as
    function of field for various $\phi/\pi$ values in the NAF sector
     ($\phi$- values correspond to curve sequence for h = 0.)
    Note the divergence close to $h_s$ for $\phi$ approaching the
    sector of the stacked dimer phase.  In this regime the zero field
    susceptibility tends to zero indicating the breakdown of the (1/S) approximation. 
    Right panel: Zero-field normalized susceptibility $\chi_n$ in the $\phi > 0$ part of the NAF 
    sector. At $\phi=0$ (see also left panel) the first order $\chi_n$ (dashed line) obtained from
    Eqs.~(\ref{NORMSUS}) and (\ref{NORMSUSAB})   
     is already reduced to half the classical value (equal to S). It decreases further with increasing
    $\phi$ and then becomes negative, indicating the instability of the NAF state. Inclusion of second order
     contributions in (1/S) stabilizes the positive value (full line, symbols)  but they diverge on approaching 
     the CAF/NAF boundary around $\phi/\pi\simeq$ 0.15. }
    \label{fig:chiNAF}
\end{figure*}
%
The zero temperature magnetization is given by the field derivative
of the total ground state energy:
\bea
m&=&m_0+m_{zp}=
-\frac{\partial E_0(h)}{\partial h} 
-\frac{\partial E_{zp}(h)}{\partial h} \ , \no\\
m&=&S\frac{h}{h_s}-\frac{1}{2N}\sum_\bk\frac{\partial\epsilon_\bk(h)}{\partial h}\ ,
\label{MAGNETIS}
\eea
where the first term is the linear classical part and the second one
the (negative) correction due to quantum 
fluctuations included up to first order in $1/S$. We can write
\bea
\frac{\partial\epsilon_\bk(h)}{\partial
h}&=&\frac{1}{\epsilon_\bk(h)}\frac{2C_\bk}{h}[(A_\bk +
C_\bk)-B_\bk]\no\\
&=&\frac{2S}{h_s}c_\bk\Bigl(\frac{a_\bk+c_\bk}{a_\bk+c_\bk\cos\theta_c}\Bigr)^\frac{1}{2}\cos\frac{\theta_c}{2} \ .
\label{DERIV1}
\eea
This finally leads to a total magnetization, including the first order quantum
corrections:
\bea
m = S\, \frac{h}{h_s}\:\Bigl[1-\frac{1}{h_s}\frac{1}{N}
\sum_\bk
c_\bk\bigl(\frac{a_\bk+c_\bk}{a_\bk+c_\bk\cos\theta_c}\bigr)^\frac{1}{2}\Bigr] \ ,
\label{MAGGEN}
\eea
where on the r.h.s. the classical value of $\theta_c$ given by
$\cos(\theta_c/2)=h/h_s$ has to be used. Because
$h_s=2S(a_\parallel +a_\perp)$ the second term in Eq.~(\ref{MAGGEN}) is
formally  a $1/S$ correction to the linear classical term
$m_0=S(h/h_s)$. 

Explicity, using Eqs.~(\ref{COEFF2},\ref{GSTRUC}) we
have for the NAF case a magnetization depending on field strength and
frustration angle according to
\begin{widetext}
\bea
\mbox{NAF}: \;\; m = S\,\frac{h}{h_s} \Bigl[1-\frac{1}{2SN}
\sum_\bk\gamma_\bk\Bigl(\frac{1+\gamma_\bk-j(1-\barg_\bk)}
{1+\gamma_\bk\cos\theta_c-j(1-\barg_\bk)}\Bigr)^\frac{1}{2}\Bigr] \ ,
\label{MAGNAF}
\eea
\end{widetext}
where we used $j=\tan\phi=J_2/J_1$. In this expression the
$(1/S)$ character of the quantum correction becomes manifest. For the
simple NAF with $j=0$ we reproduce the result first given in 
Ref.~\onlinecite{Zhitomirsky98}.

A similar but more complicated expression may be given for the CAF
phase. Defining $\delta_\bk=(1/2)(\cos k_x -\cos k_y)$ in addition to
Eq.~(\ref{GSTRUC}) we obtain
\begin{widetext}
\bea
\mbox{CAF}: \;\; m = S \frac{h}{h_s} \Bigl[1-\frac{1}{2SN(j+\frac{1}{2})}
\sum_\bk[\frac{1}{2}(\gamma_\bk +\delta_\bk)+j\barg_\bk]
\Bigl(\frac{j(1+\barg_\bk)+\gamma_\bk}
{j(1+\barg_\bk\cos\theta_c)+\gamma_\bk\cos^2\frac{\theta_c}{2}-\delta_\bk\sin^2\frac{\theta_c}{2}
}\Bigr)^\frac{1}{2}\Bigr] \ .
\label{MAGCAF}
\eea
\end{widetext}

Note that the special CAF case with $J_1=0$ ($j=\infty$) is equivalent to the simple
NAF case $J_2$ = 0 ($j=0$) in Eq.~(\ref{MAGNAF}). This may be seen by applying a
$\bk$-coordinate rotation by $\pi/4$ in Eq.~(\ref{MAGCAF}).

Naturally, the above expressions should be primarily valid deep inside
the NAF and CAF regions where the staggered moments are large. Close to
the boundaries quantum fluctuations grow and destroy the magnetic
order, then  corrections to $m(h)$ starting from the ordered
state and expanded in orders of $1/S$ are no longer appropriate. 

The combined analytical spin wave and numerical Lanczos results for the 
magnetization are shown in Fig.~\ref{fig:BonnerFisherZM1}. One may indeed see
that the agreement of both is  good deep inside the NAF ($\phi /\pi = -0.21, -0.17$) 
and  CAF ($\phi /\pi = 0.41$) sectors. On the other hand close to the classical phase 
boundaries ($\phi/\pi = 0.17,0.75$) discrepancies appear. For the former case 
our and previous\cite{Honecker01} Lanczos results
indicate appearance of the one-half magnetization plateau with $m=\frac{1}{2}m_\text{sat}$
close to $h/h_s = 0.5$. (Such a plateau has been also found in the equivalent classical
model at finite temperatures.\cite{Zhitomirsky00}
For the latter the magnetization curve becomes very nonlinear due 
to quantum fluctuations and $m(h)$ becomes negative at low fields 
(upper dashed line) indicating the breakdown of the $1/S$ expansion. 
This means that the CAF state becomes unstable and a new 
(spin nematic) order parameter will be realized in a finite sector around 
$\phi/\pi = 0.85$ ($J_2/J_1 \sim -0.5$). 
Generally, one may say that quantum corrections 
leading to nonlinear magnetization will be considerable larger on the
ferromagnetic ($J_1<0$) side of the CAF sector. This will be further 
discussed in Sec.~\ref{sect:DISC}.

\subsection{First order quantum corrections to the magnetic susceptibility}
\label{subsect:SUSC}

%
\begin{figure*}
    \includegraphics[clip,width=9.5cm]{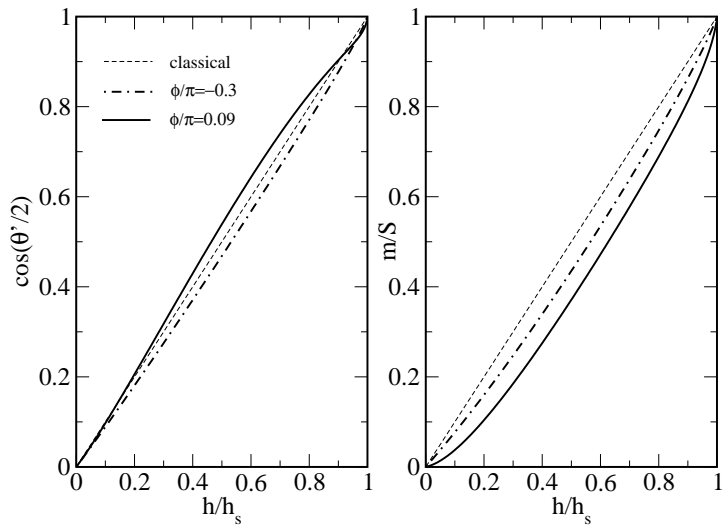}\hfill
    \caption{Comparison of (1/S) quantum corrected
    canting angle $\theta'_c$ (left) and magnetization (right) for two
    values in the NAF sector.  It is seen that first order quantum corrections
    always reduce the magnetization with respect to the classical
    value according to Eq.~(\ref{MAGGEN}).  For
    $\cos(\theta'_c/2)$ however the corrections may have both signs.
    They are small and negative for $\phi$ deep in the NAF sector
    while they are positive closer to the NAF/CAF boundary.}
    \label{fig:tilt-mag}
\end{figure*}
%
In the spin-wave approximation the magnetization curve for the simple AF
exhibits a logarithmic singularity of the slope close to the
saturation field $h_s$ as shown in Refs.~\onlinecite{Gluzman93,Zhitomirsky98}.
This should be more easily visible in the high field susceptibility.
Furthermore for $\phi$ approaching a classical phase boundary the low
field susceptibility has to vanish, which signifies the instability of
the order parameter. For these reasons we found it useful to study
the magnetic susceptibility $\chi(h,\phi)=\partial m(h,\phi)/\partial
h$ as function of the frustration angle $\phi$ in addition to the
magnetization. From Eq.~(\ref{MAGNETIS}) we obtain
\bea
\chi(h,\phi)=\chi_0+\chi_{zp}=\frac{S}{h_s} 
-\frac{1}{2N}\sum_k\Bigl(\frac{\partial^2\epsilon_\bk(h)}{\partial
h^2}\Bigr) \ .
\label{SUSDEF}
\eea
From Eq.~(\ref{DISP2}) and using $(\partial A_\bk/\partial h) = 0$ we
arrive at
\bea
\frac{\partial^2\epsilon_\bk(h)}{\partial
h^2}=\Bigl(\frac{\partial\epsilon_\bk}{\partial h}\Bigr)
\Bigl[\frac{1}{h}-\frac{1}{\epsilon_\bk(h)}\Bigl(\frac{\partial\epsilon_\bk(h)}{\partial
h} \Bigr)\Bigr] \ .
\label{DERIV2}
\eea
Inserting this in Eq.~(\ref{SUSDEF}) and using Eq.~(\ref{DERIV1}) leads to a general expression of the
normalized susceptibility $\chi_n(h)=h_s\chi(h)$ according to
\bea
\chi_n(h,\phi)=S\Bigl[1-\frac{1}{S}\bigl(\Delta\chi_n^{(a)}(h,\phi)-\Delta\chi_n^{(b)}(h,\phi)\bigr)\Bigr] \ ,
\label{NORMSUS}
\eea
where the terms $\sim 1/S$ in brackets are the quantum corrections of $\chi_n^{zp}$
to the constant classical value $\chi_n^0= S$. We obtain from Eqs.~(\ref{SUSDEF},\ref{DERIV2}): 
\bea
\Delta\chi_n^{(a)}(h,\phi)&=&\frac{S}{h_s}\frac{1}{N}
\sum_\bk c_\bk\Bigl(\frac{a_\bk+c_\bk}{a_\bk+c_\bk\cos\theta_c}\Bigr)^\frac{1}{2} \ , \\
\Delta\chi_n^{(b)}(h,\phi)&=&2\frac{S}{h_s}\Bigl(\frac{h}{h_s}\Bigr)^2\frac{1}{N}
\sum_\bk c_\bk^2\frac{(a_\bk+c_\bk)^\frac{1}{2}}{(a_\bk+c_\bk\cos\theta_c)^\frac{3}{2}} \ .
\nonumber
\label{NORMSUSAB}
\eea
For the NAF case this may be explicitly evaluated as
\begin{widetext}
\bea
\Delta\chi_n^{(a)}(h,\phi) & = & \frac{1}{2N}
\sum_\bk \gamma_\bk\Bigl(\frac{1+\gamma_\bk-j(1-\barg_\bk)}
{1+\gamma_\bk\cos\theta_c-j(1-\barg_\bk)}\Bigr)^\frac{1}{2} \ , \no\\
\Delta\chi_n^{(b)}(h,\phi)&=&\Bigl(\frac{h}{h_s}\Bigr)^2\frac{1}{N}
\sum_\bk
\gamma_\bk^2\frac{(1+\gamma_\bk-j(1-\barg_\bk))^\frac{1}{2}}
{(1+\gamma_\bk\cos\theta_c-j(1-\barg_\bk))^\frac{3}{2}} \ .
\label{SNAFAB}
\eea
\end{widetext}
Obviously only the first part  $\Delta\chi_n^{(a)}$ contributes to the (1/S) corrections of the zero-field
susceptibility. It may also be directly obtained by differentiation of m(h) in Eq.~(\ref{MAGNAF}).
For the CAF case similar expressions for 
$\Delta\chi_n^{(a,b)}(h,\phi)$ may be derived by making analogous
substitutions in the integrals and their prefactors
as done in Eqs.~(\ref{MAGNAF}) and (\ref{MAGCAF}).

The typical field dependence of the susceptibility is shown in 
Fig.~\ref{fig:chiNAF}. For $\phi/\pi = -0.49$ close to the NAF/FM boundary one
obtains nearly the classical constant value $\chi_n$ = S because in the FM
sector quantum fluctuations are not present, they are gradually turned on when
$J_1$ becomes positive and $\phi$ moves into the NAF sector. This can be clearly
seen from the various curves in Fig.~\ref{fig:chiNAF} (left). For an angle 
$\phi/\pi=0.1$ frustration becomes large and quantum fluctuations are close 
to destroying the NAF order. Accordingly the zero-field
susceptibility is close to becoming negative where the spin-wave theory breaks down. 
One also notes the upturn in the susceptibility just below 
the critical field coming from the logarithmic
singularity of the magnetization. \cite{Gluzman93,Zhitomirsky98}
The singularity becomes stronger when the strongly frustrated point
$j=1/2$ is approached. \cite{Jackeli04}
 
\subsection{Quantum corrections of the moment canting angle}
\label{subsect:CANT}

The angle $\theta_c$ between canted AF ordered moments has sofar been
given in classical approximation from minimizing only
$E_0(\theta_c,h)$. Its quantum corrections of first order in (1/S) may be computed by
minimizing the total energy
 $E_0(\theta_c,h)+E_{zp}(\theta_c,h)$, keeping $\theta_c$ also as
variable in the zero point energy. For the unfrustrated N\'eel AF
this correction is small, however, as we shall see in the following
it may be of considerable size close to the classical boundaries of
NAF and CAF phases where frustration effects are large.

The equilibrium condition including quantum corrections is given by
\bea
\frac{\partial E_0}{\partial \theta_c}+\frac{1}{2N}\sum_\bk
\bigl(\frac{\partial\epsilon_\bk}{\partial\theta_c}-\frac{A_\bk}{\partial\theta_c}\bigr)
= 0 \ .
\label{EQUILZP}
\eea
Now $\theta_c$ is to be treated as a variable present in
$\epsilon_\bk$ and $A_\bk$. The form of $A_\bk$ in
Eqs.~(\ref{COEFF1},\ref{COEFF2})  has already the classical angle of
Eq.~(\ref{CLASSANG}) substituted. 
 Its general form is $A_\bk=Sa_\bk$ with
\bea
a_\bk&=&-4J_2(1-\barg_\bk)-4J_1\cos\theta_c+h\cos\frac{\theta_c}{2} \ ,
\label{NAFCOEFF3}
\eea
for the NAF case. Likewise for CAF sector one obtains
\bea
a_\bk&=&-2J_1(1-\gamma_y)\nonumber\\
&&-2(J_1+2J_2)\cos\theta_c+h\cos\frac{\theta_c}{2} \ .
\label{CAFCOEFF3}
\eea
Evaluating the derivatives of $\epsilon_\bk$ and $A_\bk$
with respect to $\theta_c$ and solving the equilibrium
Eq.~(\ref{EQUILZP}) we obtain the canting angle $\theta'_c$
renormalized by quantum fluctuations:
\begin{widetext}
\bea
\cos\frac{\theta'_c}{2}=\cos\frac{\theta_c}{2}
\Bigl\{1-\frac{1}{h_s}\frac{1}{N}\sum_\bk
\Bigl[\frac{a_\bk^{(1)}(a_\bk+c_\bk\cos^2\frac{\theta_c}{2})+c_\bk(a_\bk+c_\bk)}
{(a_\bk+c_\bk)^\frac{1}{2}(a_\bk+c_\bk\cos\theta_c)^\frac{1}{2}}
-a_\bk^{(1)}\Bigr]\Bigr\} \ .
\label{TILTGEN}
\eea
\end{widetext}
Here $\cos\frac{\theta_c}{2}=h/h_s$ is the classical canting angle and 
$a_{\bf k}^{(1)}=-(h_s/2S)$ is the coefficient of the second 
term in Eqs.~(\ref{NAFCOEFF3},\ref{CAFCOEFF3}).
Again, because of $h_s=2S(a_\parallel+a_\perp)$ the term 
$\sim 1/h_s$ in Eq.~(\ref{TILTGEN}) has to
be considered as a (1/S) correction to the canting angle. Therefore
in the sum on the r.h.s.  the classical values for $\theta_c$ has to
be used which leads again to the intra-sublattice coupling $a_\bk$ as
given by Eq.~(\ref{COEFF2}). 

The general solution in
Eq.~(\ref{TILTGEN}) is valid for both AF phases. In the NAF
case it leads to the explicit result
\begin{widetext}
\bea
\cos\frac{\theta'_c}{2}=\cos\frac{\theta_c}{2}
\Bigl\{1-\frac{1}{2SN}\sum_\bk
\Bigl[\frac{\gamma_\bk^2+\gamma_\bk\sin^2\frac{\theta_c}{2}-1+j(1-\barg_\bk)(1-\gamma_\bk)}
{(1+\gamma_\bk-j(1-\barg_\bk))^\frac{1}{2}(1+\gamma_\bk\cos\theta_c-j(1-\barg_\bk))^\frac{1}{2}}
+1\Bigr]\Bigr\} \ .
\label{TILTNAF}
\eea
\end{widetext}
For the simple AF ($j=0$) one recovers the expression given in
Ref.~\onlinecite{Zhitomirsky98}. A similar explicit expression for
the CAF case may be derived but it is too unwieldy to be given here.
For the numerical calculation of $\cos\frac{\theta'_c}{2}$
Eq.~(\ref{TILTGEN}) may be used as well.

The field dependence of renormalized moment canting given 
by $\cos\frac{\theta'_c}{2}$ in comparison with the normalized magnetization m(h)/S is shown in
Fig.~\ref{fig:tilt-mag}. Generally the quantum corrections to the canting angle are
quite small deep inside the AF sectors. As for the magnetization they become larger
when approaching a phase boundary. Interestingly however, they may have different signs for 
the former (left panel) while they must always be negative for the latter (right panel).
The positive correction to  $\cos\frac{\theta'_c}{2}$ appears in the vicinity of the NAF/CAF boundary.

The quantum corrections to magnetization and canting angle may be
used to obtain the correction to the moment size.
Classically we have  S =
$m_0/\cos\frac{\theta_c}{2}$. If we use a similar definition
including the quantum
corrections we  have 
 $\langle S\rangle=m/\cos\frac{\theta'_c}{2}$ for
the renormalized moment. The change of moment 
size defined by $\delta S = S- \langle S\rangle$ is then given by
$\delta S=S-(m/\cos\frac{\theta'_c}{2})$. Note however
that in this relation the quantum correction to S formally contains
effects of arbitrary order in (1/S) even if m and
$\cos\frac{\theta'_c}{2}$ are only corrected in order (1/S). Close to
the nonmagnetic regions  when m aproaches zero $\delta S/S$ becomes
unity, i.e., the staggered moment is destroyed by the quantum
fluctuations.

\subsection{Second-order spin-wave results for the magnetization and susceptibility}
\label{subsect:MAGNET2}
%
\begin{figure*}
\includegraphics[width=0.5\textwidth]{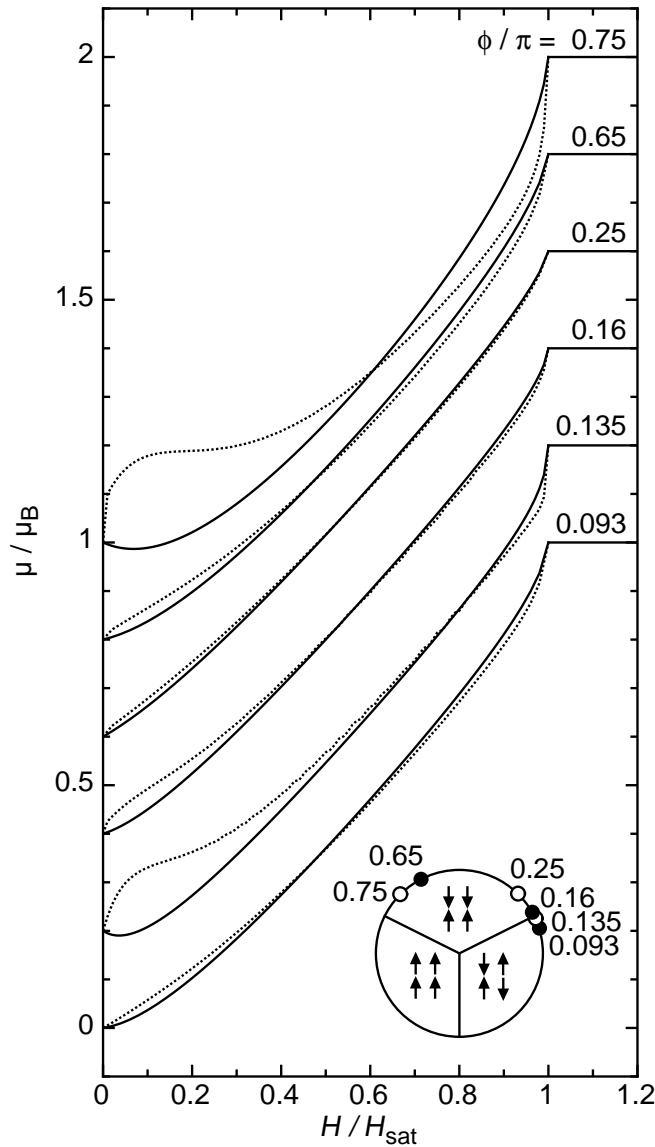}
\caption{Compiled comparison of first (full lines)  and second (dotted lines) order in (1/S) spin wave results for the magnetization for various frustration  angles indicated in the inset. An offset of 0.2 has been applied. Curves in increasing order correspond to $\phi$- values in counter-clockwise direction. Well inside the antiferromagnetic regimes ($\phi/\pi = 0.093, 0.25, 0.65$) first and second order results show little difference. Close to the classical phase boundaries ($\phi/\pi = 0.135, 0.16, 0.75$) first order results lead to instable (negative) low-field magnetization. The instability region below $\phi/\pi$ = 0.85 (CAF/FM) is much larger than around $\phi/\pi$ = 0.15 (NAF/CAF). Second order results are are always positive but show very anomalous low field magnetization  contrary to the Lanczos results in Fig.~\ref{fig:BonnerFisherZM1}. Note that the second order corrections in (1/S) to the first order curves are positive for low fields and negative for large fields.}
\label{fig:magn_12}
\end{figure*}
%
The linear spin wave theory used in the previous sections
includes the leading contribution to the ground state energy 
$\sim {\cal O}(S)$, see Eq.~(\ref{HPHAM}).
The next order contributions to $E_{\rm g.s.}$ come 
(i) from the canting angle renormalization, (ii) from a Hartree-Fock 
decoupling of the quartic terms, and (iii) from the cubic terms.
The detailed derivation in the case of nonfrustrated square
lattice AF is given in Ref.~\onlinecite{Zhitomirsky98}.
Here, we present only the final result for the N\'eel phase
of the $J_1$--$J_2$ model.
The ground state energy per site in second order spin-wave theory
consists of the classical part $E_0$, first order corrections (zero 
point fluctuations) $E_{zp}$ and second order in $1/S$ corrections.
In total it may be written as
\begin{widetext}
\begin{eqnarray}
E_{\rm g.s.}/N & = & -2(J_1-J_2)S(S+1) - \frac{h^2}{16J_1} +
\frac{1}{2N} \sum_{\bf k} \varepsilon_{\bf k}\nonumber\\
&&\mbox{}
-2J_1 \bigl[(n-\delta_1)^2 + n_1(n_1-\delta)\bigr]
+ 2J_2 \bigl[(n-n_2)^2 + \delta_2(\delta_2-\delta)\bigr]\\
& & \mbox{}
+  4J_1\cos^2(\theta_c/2)\Bigl[(n-\frac{1}{2}\delta)(\delta_1+n_1)
-2\delta_1n_1\Bigr]
-J_1\sin^2\theta_c \frac{1}{3N^2} \sum_{\bf k,q}
\frac{F({\bf k},{\bf q})^2}
{\varepsilon_{\bf k} + \varepsilon_{\bf q} +\varepsilon_{\bf k+q-Q}} \ .
\nonumber
\label{Egs2}
\end{eqnarray}
\end{widetext}
Here the first line is equal to the sum of $E_0+E_{zp}$ while the second 
and third line constitute the second order in $1/S$ spin-wave contribution.
In this term the six constants are given by two-dimensional momentum integrals:
\begin{eqnarray}
&& n = \frac{1}{N} \sum_{\bf k} \frac{A_{\bf k}+C_{\bf k} -\varepsilon_{\bf k}}
{2\varepsilon_{\bf k}} \ , 
\nonumber \\
&& n_{1(2)} = \frac{1}{N} \sum_{\bf k}\gamma_{\bf k}(\bar{\gamma}_{\bf k})\,
\frac{A_{\bf k}+C_{\bf k}}{2\varepsilon_{\bf k}}\ , \\ 
&& \delta = \frac{1}{N} \sum_{\bf k} \frac{B_{\bf k}} {2\varepsilon_{\bf k}} \ , \ \ \
\delta_{1(2)} = \frac{1}{N} \sum_{\bf k} \gamma_{\bf k}(\bar{\gamma}_{\bf k})\,
\frac{B_{\bf k}}{2\varepsilon_{\bf k}} \ , \nonumber
\end{eqnarray}
while the expression for the cubic vertex  $F({\bf k},{\bf q})$
is given by Eq.~(25) in Ref.~\onlinecite{Zhitomirsky98}.
The magnetization curve is obtained by numerical differentiation:
$m = - \partial E_{\rm g.s.}(h)/\partial h$.

The effect of the second order contributions in the magnetization may be seen
in Fig.~\ref{fig:magn_12} and also in Fig.~\ref{fig:BonnerFisherZM1} 
in the comparison to the ED Lanczos results. In general the second order 
spin-wave corrections to the first-order in $(1/S)$ results are positve for small
fields and negative for large fields (Fig.~\ref{fig:magn_12}). As long as $\phi$ is within the stable 
antiferromagnetic sectors the second order corrections are quite small. 
For example when $\phi/\pi$ = -0.21, -0.17, 0.41 first- and second-order results 
are almost indistinguishable to the eye and agree with Lanczos results (Fig.~\ref{fig:BonnerFisherZM1}).
When $\phi$ approaches the strongly frustrated phase boundaries first-order approximation 
breaks down as witnessed by the magnetization becoming negative, 
Figs.~\ref{fig:BonnerFisherZM1} and \ref{fig:magn_12}. 
The second-order corrections remedy this situation.  However, close to the boundaries they 
become very large leading to a very anomalous low field second order magnetization 
which deviates strongly from the Lanczos results (Fig.~\ref{fig:BonnerFisherZM1}). 
Since the deviations between first and second  order curves
are strongly enhanced close to the classical phase boundaries  it is clear that spin wave 
expansion no longer converges.
In  Fig.~\ref{fig:BonnerFisherZM1} this is obvious for  $\phi/\pi$ = 0.17 at the NAF/CAF boundary and even more for 
 $\phi/\pi$ = 0.75, already well in advance of the CAF/FM boundary at $\phi/\pi$ = 0.85.
 In fact the first order result (long-dashed) agrees better with the Lanczos results (full symbols)  
plotted for comparison than does the second order curve (short-dashed).

Finally, we mention that from Eq.~(\ref{Egs2}) the  second order in $(1/S)$ corrections 
to the zero-field susceptibility may be calculated in addition to the first order 
expression given in Eqs.~(\ref{NORMSUS}) and (\ref{NORMSUSAB}). The comparison of 
first and second order results for the NAF case is given in Fig.~\ref{fig:chiNAF} 
(right panel). Although the second order contributions repair the negative 
instability of the first order susceptibility at $\phi\simeq 0.11\pi$ the second 
order result itself diverges when one moves even closer to the classical phase 
boundary at $\phi\simeq 0.15\pi$.\\

\section{Effect of interlayer exchange coupling}
\label{sect:interlayer}

In real magnetic systems other interactions may play
a certain role besides the in-plane Heisenberg exchange.
These include various anisotropies as well as a three-dimensional
coupling.
Below we shall consider modifications of the above formulas produced
by interlayer exchange coupling:
\begin{equation}
{\cal H}_\perp = J_\perp \sum_{\langle ij\rangle_z} {\bf S}_i\cdot{\bf S}_j \ ,
\end{equation}
for nearest-neighbor spins in the direction perpendicular  to the layers. (Note the different convention
for field direction in Sect.~\ref{subsect:SW}).
Here we assume the simple stacking of the layers. The obtained results can be easily
extended to other cases as well.

For ferromagnetic exchange $J_\perp<0$, the saturation
field for NAF is still given by $h_s = 8SJ_1$, while for  antiferromagnetic exchange 
$J_\perp>0$ it is given by  $h_s=8SJ_1(1+\frac{1}{2}j_\perp)$. Here, we define 
$j_{\perp} =J_\perp/J_1$. The first-order spin-wave result for 
the magnetization in the NAF case which includes the interplanar coupling 
$J_{\perp}$ is given by 
\begin{widetext}
\begin{eqnarray}
J_\perp < 0: \;\; m &=& S \frac{h}{h_s}\biggl[ 1 - \frac{1}{2SN} \sum_{\bf k} \gamma_{\bf k} 
\Bigl( \frac{ 1+\gamma_{\bf k}-j (1-\bar{\gamma}_{\bf k})
+ \frac{1}{2}|j_\perp|(1-\cos k_z)}
{1+\gamma_{\bf k}\cos\theta_c-j(1-\bar{\gamma}_{\bf k})   
+ \frac{1}{2}|j_\perp|(1-\cos k_z)}\Bigr)^{1/2} \biggr],\\
J_\perp>0:\;\; m &=& S \frac{h}{h_s}\biggl[ 1 - \frac{1}{2SN} \sum_{\bf k} 
\frac{\eta_{\bf k}}{1+\frac{1}{2}j_\perp} 
\Bigl(\frac{1+\frac{1}{2}j_\perp +\eta_{\bf k}-j(1-\bar{\gamma}_{\bf k})}
{1+\frac{1}{2}j_\perp+\eta_{\bf k}\cos\theta_c-j(1-\bar{\gamma}_{\bf k})}\Bigr)^{1/2}  
\biggr]\ .\nonumber
\end{eqnarray}
\end{widetext}
Here $\eta_{\bf k} = \gamma_{\bf k} + \frac{1}{2}j_\perp\cos k_z$ is the 3D structure factor and 
summation is now extended over a 3D Brillouin zone.

In the case of CAF order the  first-order spin-wave
theory for a ferromagnetic interlayer exchange $J_\perp<0$ leads to 
$h_s=4SJ_1(1+2j)$  and for antiferromagnetic case  $J_\perp>0$ 
the saturation field is given by  $h_s = 4SJ_1(1+2j+4j_\perp)$. 
The first order spin wave result for the magnetization in the 3D case is obtained as
\begin{widetext}
\begin{eqnarray}
J_\perp<0: \;\; m &=& S\frac{h}{h_s}\biggl[ 1 - \frac{1}{2SN} \sum_{\bf k} 
\frac{j\bar{\gamma}_{\bf k}+\frac{1}{2}\gamma_x}{j+\frac{1}{2}} 
\Bigl(\frac{j(1+\bar{\gamma}_{\bf k}) + \gamma_{\bf k} + \frac{1}{2}|j_\perp|(1-\gamma_z)}
{j(1+\bar{\gamma}_{\bf k}\cos\theta_c) + \frac{1}{2} (\gamma_y+\cos\theta_c \gamma_x)   
+ \frac{1}{2}|j_\perp|(1-\gamma_z)}\Bigr)^{1/2} \biggr], \\
J_\perp<0: \;\; m &=& S \frac{h}{h_s}\biggl[ 1 - \frac{1}{2SN} \sum_{\bf k} 
\frac{j\bar{\gamma}_{\bf k}+\frac{1}{2}\gamma_x +\frac{1}{2}j_\perp \gamma_z}{j+\frac{1}{2}(1+j_\perp)} 
\Bigl(\frac{j(1+\bar{\gamma}_{\bf k}) + \gamma_{\bf k} + \frac{1}{2}j_\perp(1+\gamma_z)}
{j(1+\bar{\gamma}_{\bf k}\cos\theta_c) + \frac{1}{2} (\gamma_y+\cos\theta_c \gamma_x)  
+ \frac{1}{2}j_\perp(1+\gamma_z\cos\theta_c)}\Bigr)^{1/2}
\biggr]\ .\nonumber 
\end{eqnarray}
\end{widetext}
Here we used the convention  $\gamma_{i} = \cos k_{i}$ ($i = x,y,z$).

We have verified that for well ordered NAF and CAF phases the effect of interlayer
coupling is hardly visible up to $J_\perp \sim 0.3J_{1,2}$. This means 
that the application of our first order spin wave results to real compounds 
does not critically depend on a small interlayer coupling $J_\perp$
as long as the quantum antiferromagnet is in a well-ordered phase.
Closer to the phase boundaries $J_\perp$ has the the effect of stabilizing the ordered
phases, especially for J$_\perp < 0$.

\section{Discussion and conclusion}
\label{sect:DISC}

%
\begin{figure*}[tbc]
\centering
\hfill
\includegraphics[width=7cm]{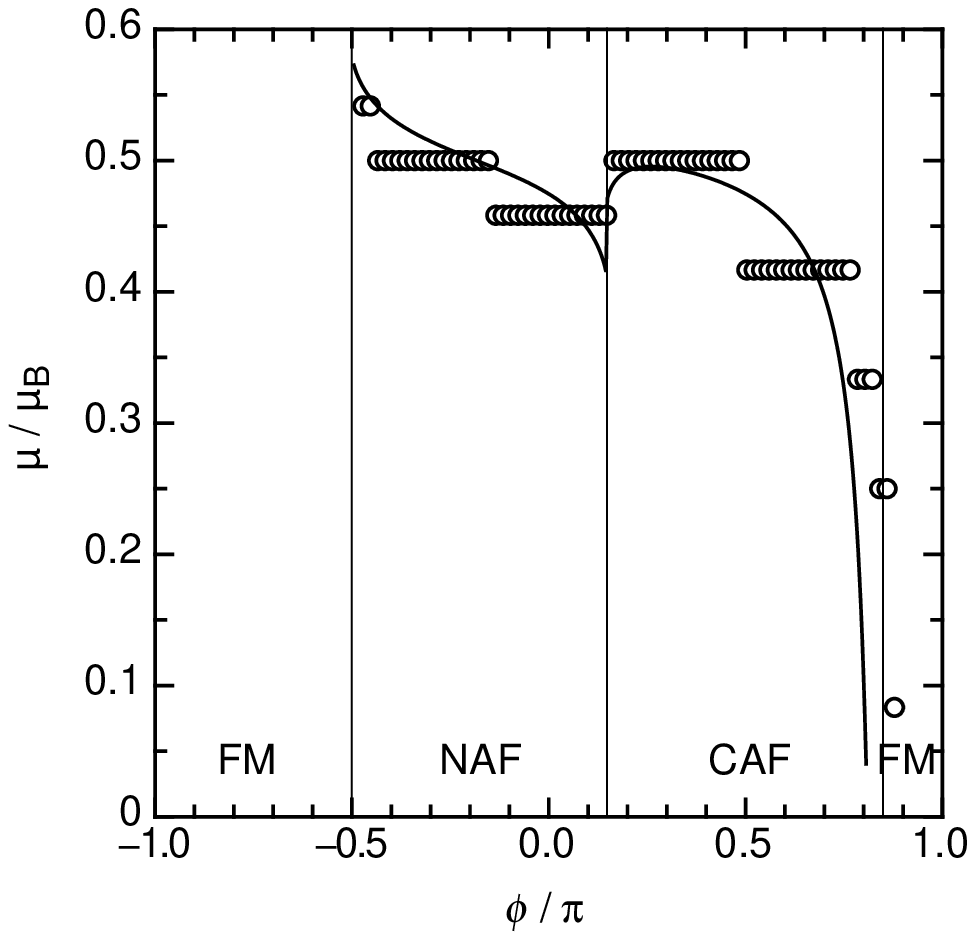}\hfill\hfill
\includegraphics[width=7cm]{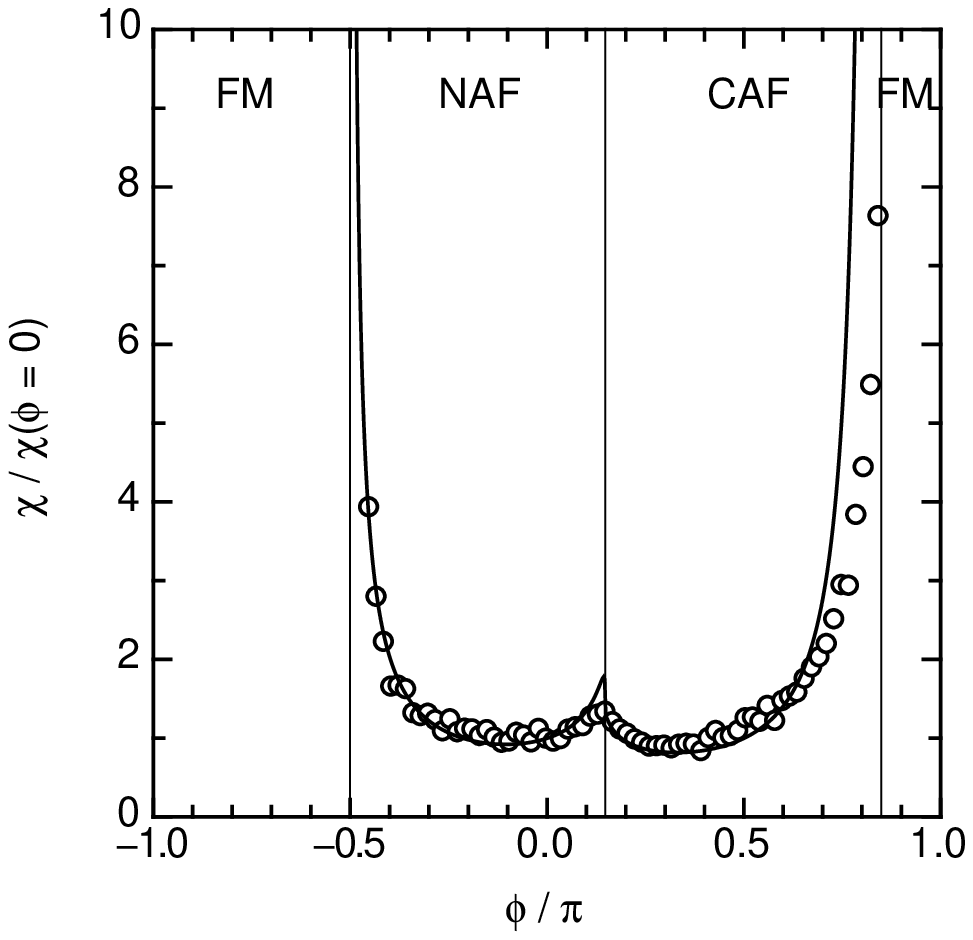}
\hfill\null
\caption{Left: Variation of the moment $\mu/\mu_B = gm$ as function of
$\phi$ for h/h$_s$=0.58 
close to half-saturation. Circles are obtained from the Bonner-Fisher
plots based on numerical (T=0) Lanczos results.
Full lines include the (1/S) quantum correction from spin wave excitations
according to Eq.~(\ref{MAGGEN}) with respect to the
constant classical value  $\mu/\mu_B = m/S = 0.58$. At the FM/NAF
boundary where the quantum fluctuations are small
the magnetization starts at this value and then is continuously
reduced as one approaches the strongly frustrated sector
around NAF/CAF boundary. There the magnetization jumps to a slightly
larger value and then breaks down close to the 
spin nematic region around CAF/FM boundary.
Right:  Susceptibility  $\chi(\phi)/\chi(\phi=0)$ for $h=h_s^-(\phi)$
normalized to the simple NAF value. Circles are ED Lanczos results (k$_B$T/J$_c$=0.2) and
full lines are T=0 spin wave results according to Eq.~(\ref{NORMSUS}).}
\label{fig:mu_at_0.58_hsat}
\end{figure*}
%

In this work we have explored the quantum corrections to the magnetization, the uniform
susceptibility and the canting angle
in the first- and the second-order spin-wave approximation and by using exact 
diagonalization of finite clusters. The deviations from classical behaviour were 
found to be pronounced close to the strongly frustrated regions where the classical 
phases meet. Indeed we have shown that linear spin-wave theory breaks down 
in these regions and second-order corrections do not fundamentally change 
this observation. The latter are small well within the AF regions and have 
positive sign for small and negative sign for large fields. Although they prevent 
the instability of the linear spin-wave theory for small fields they 
become very anomalous close to the boundaries and one cannot expect that the $1/S$ 
expansion converges in the nonmagnetic region. This is intuitively clear since 
the spin wave expansion in the region of the classical boundaries starts from 
the wrong, {\it i.e.}, magnetically ordered ground state.

A striking feature of these results is that the deviations from classical behaviour in the 
magnetization curve $m(h)$ and the breakdown of the spin-wave expansion are 
most pronounced on the ferromagnetic side ($J_1<0$), as illustrated in Figs.~2 and 5.  
One can understand this by considering the asymptotic form of the first-order
corrections to the staggered moments $\delta S$ and the uniform susceptibility
$\delta\chi$ in the CAF phase as  $j \rightarrow \pm 1/2$:
\begin{eqnarray}
\delta S & \simeq & \sum_{\bf k} \frac{1}{\sqrt{k_x^2 + |\delta j| k_y^2}}\ , \nonumber \\ 
\delta\chi & \simeq & \frac{1}{j+\frac{1}{2}}
\sum_{\bf k}\frac{1}{\sqrt{k_x^2+|\delta j| k_y^2}}\ .
\end{eqnarray}
Here, $\delta j = j\pm\frac{1}{2}$ is the deviation from one of the two
strongly frustrated points.
For the CAF/NAF boundary the diverging corrections are the same
for the sublattice magnetization and the susceptibililty:
$\delta S,\delta\chi \sim \ln|\delta j|$.
For the CAF/FM frustration point the susceptibililty correction acquires
an additional diverging prefactor:
$\delta\chi \sim (\ln|\delta j|)/(\delta j)$, which indicates that a long-range
magnetic order is destabilized in a much wider window of $J_2/J_1$
for this sign of $J_1$. 

This is in accordance with the ED Lanczos results where a tendency to 
bound state formation of spin waves as indicated by the $\Delta S_z = 2$ 
steps in the magnetisaton of finite clusters is observed\cite{Schmidt07}. 
This is evidence that around $\phi\simeq 0.85\pi$  or $J_2/J_1\simeq -0.5$ 
the ground state will be of the spin nematic type as proposed 
in Ref.~\onlinecite{Shannon06}. It may be viewed as a quantum gapless phase 
with a Goldstone mode describing the collective long range excitations of 
a nonlocal quadrupolar order parameter.    A second order transition between
and the CAF phase and this spin nematic {\it is} permitted by the symmetry of the 
order parameter, and the smooth evolution of ED spectra across the transition
suggest that a second order does in fact occur.   The pronounced quantum 
fluctuations seen in spin wave theory lend further support to this idea.             

This can be most clearly seen in Fig.~\ref{fig:mu_at_0.58_hsat} (left panel) 
where the magnetization is plotted as function of the frustration angle $\phi$. 
The decrease in magnetization from the classical value $m/S=h/h_c$ at around 
half saturation ($h/h_c = 0.58$) characterizes the strength of quantum fluctuations. 
Their effect increases from zero at FM/NAF boundary to a maximum at NAF/CAF 
boundary where a discontinuous jump in the magnetization occurs. 
For $\phi/\pi>0.5$ (the CAF regime with $J_2<0$) the reduction of $m/S$ due 
to quantum fluctuations rapidly becomes large and the expansion in $1/S$ 
breaks down.   This is also seen in Fig.~\ref{fig:magn_12}.

The exact numerical results (circles) obtained from the Bonner-Fisher plots (hence the steps) as in Fig.~\ref{fig:mu_ed} also show the strong reduction of the magnetization close to classical CAF/FM boundary . This is a signature of the true spin-nematic quantum ground state which does not break time reversal symmetry. Therefore it has no first order (linear) coupling to the magnetic field, resulting in a small magnetization. The right panel of  Fig.~\ref{fig:mu_at_0.58_hsat} shows the comparison of high-field susceptibility at $h=h^-_s(\phi)$ from first order spin wave theory (full line) at T=0 with the ED Lanczos results at small but finite T, plotted as function ot the frustration angle $\phi$. The susceptibility in both cases is normalized to the pure NAF case ($\phi$=0). The overall agreement of $\phi$ dependence is quite good because due to high fields the effect of quantum fluctuations is suppressed. Again the deviations are strongest close to the classical CAF/FM boundary where the saturation field approaches zero and the spin wave approximation breaks down.

The deviations from classical results and the breakdown of spin wave approximation 
is much less severe on the CAF/NAF phase boundary at the AF side ($J_1>0$). 
The magnetization behaves significantly less singular in this region. 
As mentioned before a discontinuous jump in $m(\phi)$ at $\phi\simeq 0.15\pi$ 
occurs around half saturation (Fig.~\ref{fig:mu_at_0.58_hsat}). 
Then the quantum phase transition to the presumably stacked spin dimer 
ground state may be expected to be a first order transition. 
Therefore the magnetic and nonmagnetic phases on both sides of the boundary 
will correspond to stable local minima of the free energy and fluctuations 
will not be very pronounced leading to a less singular magnetization behaviour. 
The first-order nature of transition between CAF and a nonmagnetic columnar
dimer state was noticed in early numerical work,\cite{Schulz92} 
while for the boundary with the NAF, a first-order scenario was put forward
only recently. \cite{Sirker07}
         
Finally, we comment on experimental data for
the high field magnetization of $J_1$--$J_2$ compounds. 
There are no published data for those systems mentioned in the introduction. 
However recently a new compound (CuBr)LaNb$_2$O$_7$, with a perovskite/metal halide intergrowth 
structure was synthesized~\cite{Oba06}.   This is a spin-1/2 magnetic insulator, and is reported to exhibit 
quasi-2D magnetic behavior arising from the CuBr - square lattice planes.   
It shows CAF order at a relatively large $T_N = 32$~K.

It has been suggested that the magnetism of (CuBr)LaNb$_2$O$_7$ can be described by the 
square lattice $J_1$--$J_2$ model with a frustration angle $\phi=0.73\pi$ ($J_2/J_1=-1.1$), 
which puts this compound closer to the strongly frustrated region 
of the spin nematic phase  than any other compound reported sofar. 
Our spin wave calculations predict a pronounced nonlinear magnetization 
for this parameter set, particularly at low fields.  However,  
the experimental results in Ref.~\onlinecite{Oba06} show only a modest curvature,
at high fields.   We therefore conclude that, while these materials likely  do possess
competing FM and AF interactions, they probably cannot be described by a simple square lattice 
$J_1$--$J_2$ model.

In conclusion, it is abundantly clear that classically disordered phases
of the square lattice $J_1$--$J_2$ model need and deserve 
an analysis which goes beyond the spin wave theory presented here. 
For further progress to be made in understanding the ``hidden order''
phases an approximative treatment of the broken order parameter and its 
low lying excitations is necessary.   Since both the dimer and nematic phases
are bond centered, the bond-operator method might be a useful choice.

\section{Acknowledgement} 

We are pleased to acknowledge helpful conversations with C. Geibel, Tsutomu Momoi  and Luis Seabra. 
This work was supported in part by the SFB 463 project of DFG and by EPSRC Grant (No. EP/C539974/1).



\end{document}